\documentclass{elsart}
\usepackage{epsfig}

\usepackage{bm} 
\usepackage{mathptmx} 
\newcommand{\be}{\begin{equation}}
\newcommand{\ee}{\end{equation}}
\newcommand{\bc}{\begin{center}}
\newcommand{\ec}{\end{center}}
\newcommand{\bi}{\begin{itemize}}
\newcommand{\ei}{\end{itemize}}
\newcommand{\ba}{\begin{eqnarray}}
\newcommand{\ea}{\end{eqnarray}}

\newcommand{\ignore}[1]{}
\begin{document}
\begin{frontmatter}

\title {Critical brain networks\thanksref{label1}}
\thanks[label1]{
Contribution to the Niels Bohr Summer Institute on Complexity and Criticality; to appear in a Per Bak Memorial
Issue of PHYSICA A. Work supported by NIH NINDS of USA (Grants 42660 and 35115). The hospitality and support of
the Universitat de les Illes Balears, Palma de Mallorca, Spain and discussions with Drs. A.V. Apkarian, G.
Cecchi, V. Eguiluz and M. Paczuski are also acknowledged. Special thanks to Dr. D. Plenz for discussions and for
providing Fig. 4.}
\author{Dante R. Chialvo}
\address{Department of Physiology, Northwestern University, 303 E.
Chicago Avenue, Chicago, Illinois, 60611, USA.}

\begin{abstract}
Highly correlated brain dynamics produces synchronized states with no behavioral value, while weakly correlated
dynamics prevents information flow. We discuss the idea put forward by Per Bak that the working brain stays at
an intermediate (critical) regime characterized by power-law correlations.
\end{abstract}


\end{frontmatter}
\section{Introduction}
The human brain is a large system, with no more than a hundred specialized modules with different functions. At
the smallest grain, the cerebral cortex consists of about $10^{10}$ neurons that comprise a highly
interconnected network. Each cell receives continuously a few thousands of excitatory inputs from other neurons.
One of the simplest things we do not know about the brain is how the cortex, being a mainly excitatory network,
prevents the expected explosive propagation of activity and still transmits information across areas. If the
average number of neurons activated by one neuron is too high (i.e., supercritical) this results in the massive
activation of the entire network, while if it is too low (i.e., subcritical), propagation dies out. The critical
regime is the one in which these opposing processes are balanced. It was Turing, about fifty years ago
\cite{turing}, the first to speculate in these terms, arguing that brains should be at a barely critical state.
We review here the proposal \cite{bak0} that most behaviorally relevant brain states, are associated with
dynamics which is critical in this sense. This perspective places the emphasis on understanding the brain's
large number of dynamical nonlinear degrees of freedom, and in the dynamical attractors that are expected to
emerge from the interaction of these elements. Less emphasis is given to some other aspects, including probable
``computation" properties of the emergent circuits. In a loose sense it claims that what we see as a {\it brain}
is all what it can be expected as the obligatory solutions of putting these degrees of freedom to interact with
each other. It says nothing about anything else.

The paper is dedicated to discuss recent experimental findings of critical correlations in brain dynamics. The
paper is organized as follows. The next section reminds ourselves where the brain problem is in the general
context of dynamical systems. In section 3 we discuss results from brain imaging experiments showing a broad
distribution of functional connectivity, implying that brain networks are scale-free. This is contrasted with
the known cortical connectivity. Section 4 contains recent experimental evidence at the in vitro level
indicating that cultured cortical networks are critical at the neuronal level. The paper concludes with a
comment on the ideas of our late friend Per, who foresaw many of these results.
\section{Brains in ``Dynamicsland".}
Brain activity happens in bursts, in which pauses, silence or boredom suddenly and unpredictably are followed by
brief activity. From a dynamical viewpoint brain dynamics is not different from other natural processes. Nature
is clearly non homogeneous and intermittent, the analysis of any natural object reveals an ever surprising
amount of details; there is no single relevant scale at which Nature becomes homogeneous. Complexity is this
lack of uniformity associated with the scale-free spatiotemporal feature. The driving force in this field
continues to be the effort to understand what generates complexity, and how many different dynamical mechanisms
can produce scale-free objects.
\begin{table}[h!tb]
  \centering
\begin{tabular}{|c|c|}
   \hline
   Complicated & many linear pieces + a central supervisor +
   blueprint\\
   Systems & = ``whole" (Example: tv set)\\
   \hline
Complex & many nonlinear pieces + coupling + injected
energy \\
Systems& = ``emergent properties" (Example: society) \\
   \hline
 \end{tabular}\\
\caption{Complicated systems are not complex.}
\end{table}
It is now widely recognized that, under a variety of conditions, non linear systems with many degrees of freedom
tend to evolve towards complexity and criticality \cite{bak0,btw}.  It is the interaction of many nonlinear
degrees of freedom which produces emergent dynamics we call complex. The latest ``heavenly example" is the sun's
sudden bursts of radiation emanating from quick re-arrangements of the magnetic field network in the corona
\cite{Paczuski04}. This is different from the dynamics arising in complicated systems comprised of the simple
addition of interconnected {\it linear} pieces (Table 1). Examples of complicated systems are a television set,
or a car. They do not ``emerge", they are manufactured by following a blueprint given by the designer. Evolution
to more complicated designs requires always of the supervisor (or designer) intervention. Complex systems such
as species, ecologies, societies or brains do not arise from blueprints, they are very robust emergent
consequences of dynamical laws we still do not understand.
 We can not replay the tape of evolution to investigate
whether other ecologies will arise or self-organize, but in the case of brains we witness the high percentage of
brains that end up well connected and, with few exceptions, working. This is a marvellous thing, considering
that during one year of development a brain is continuously adding (or ``connecting") $10^5$ news neurons per
minute. What are the basis of the self-organizing mechanism able to achieve such feats?
\begin{figure}[h!tb]
\centering \psfig{figure=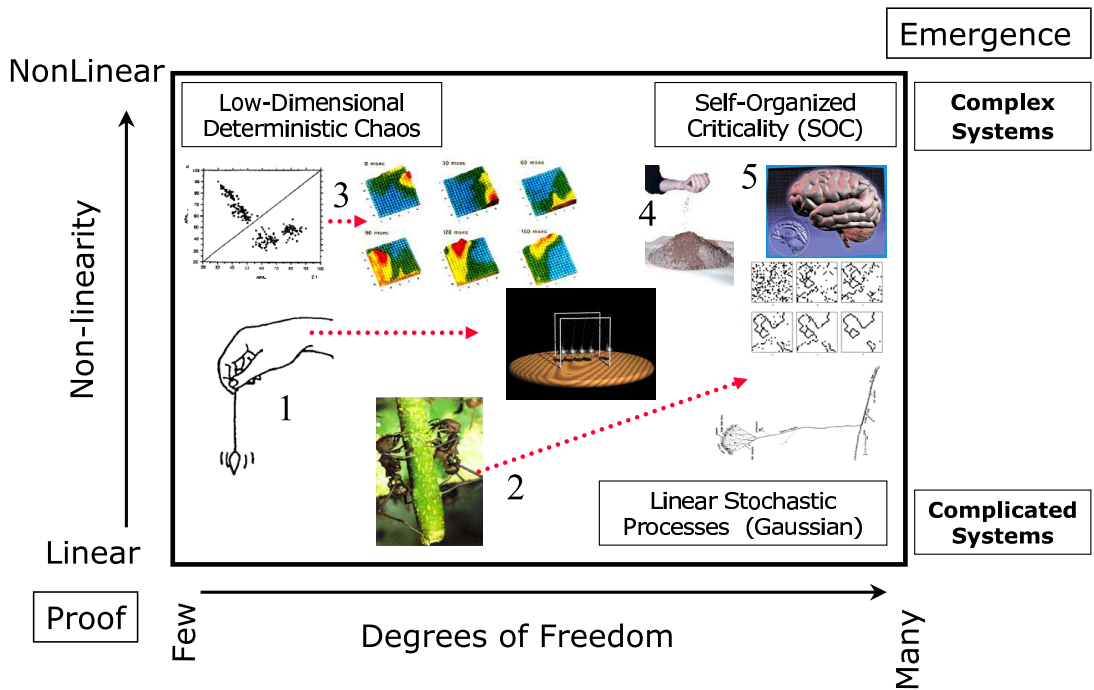,width=5.68 truein,clip=true,angle=0}
\caption{\footnotesize{``DynamicsLand": A cartoonish representation of the parameter space for various classes
of dynamical systems. The simplest ones ``live" in the left bottom corner, where analysis and formal proofs are
the techniques expected, but many fundamental problems in biology correspond to areas distant from that land.
Relatively simple dynamics gets sophisticated as the nonlinear term acquires relevance (moving upward in the
graph) or as the number of degrees of freedom increases (moving to the right). Pictorial examples include: (1)
the transition from one to many coupled pendulums, (2) few foraging ants to the entire swarm \cite{millonas},
(3) from the chaotic dynamics of an isolated cardiac cell \cite{chialvo90} to the spatiotemporal spiral waves in
the heart \cite{davidenko90}, (4) a sandpile and, of course, (5) the brain. }}
\end{figure}
One motivation for biologists to look at the physical laws governing complex systems of all kinds is the hope
that universality will give us an edge. The rationale is that a good understanding of these universal laws will
provide a breakthrough and shed light on related biological problems. The considerations in the cartoon of
Figure 1 remind us that the brain is at a region of parameter space where complex dynamics can emerge. In that
top-right corner, theory is scarce, but some insight and  numerical tools can be borrowed from related work in
self-organized criticality and complex networks, as discussed in the next section.
\section{Functional networks are scale-free.}
Brain activity is eminently spatio-temporal, as such the monitoring of the complicated cortical patterns have
greatly benefited from techniques developed in the context of functional magnetic resonance imaging (fMRI).
However, the numerical analysis of such spatiotemporal patterns is less developed, lacking mathematical tools
and approaches specifically tailored to grasp the complexity of brain cortical activity. One possibility is to
get insight from recent work showing that disparate systems can be described as complex networks, that is
assemblies of nodes and links with nontrivial topological properties \cite{albert2000,newman2003a,jasny2003}.

The brain creates and reshapes continuously complex functional networks of correlated dynamics responding to the
traffic between regions, during behavior or even at rest. We have recently studied these networks, using
functional magnetic resonance imaging in humans (see methods in \cite{eguiluz}. The data is analyzed in the
context of the current understanding of complex networks (for reviews see
\cite{albert2000,newman2003a,jasny2003,doro2002,strogatz2001,watts1998}). During any given task the networks are
constructed in the following way. Magnetic resonance brain activity is measured, at each time step (typically
400 spaced 2.5 sec.), from $36\times64\times64$ brain sites (so-called ``voxels'' of dimension $3\times
3.475\times 3.475~$mm$^3$). The activity of voxel $x$ at time $t$ is denoted as $V(x,t)$.
\begin{figure}[h!tb]
\centering \psfig{figure=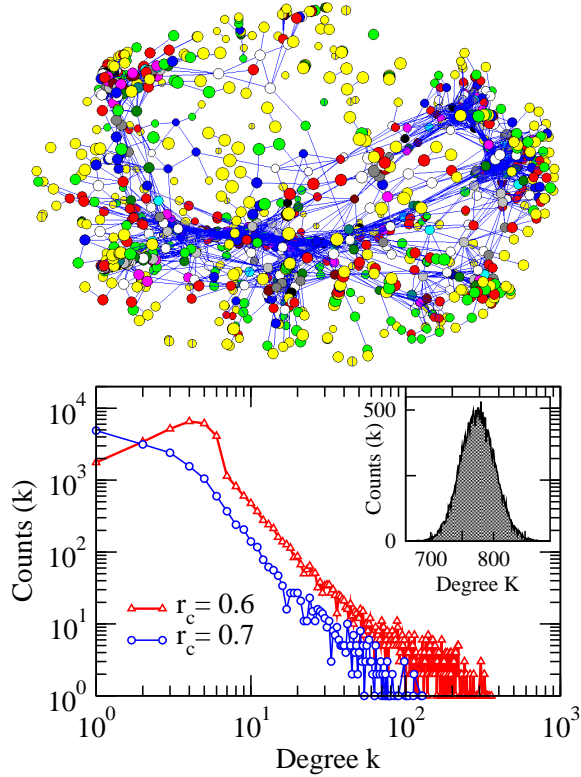,width=3 truein,clip=true,angle=0} \psfig{figure=figure2b.eps,width=3
truein,clip=true,angle=0}\caption{\footnotesize{A typical brain network extracted from functional magnetic
resonance imaging. Top panel shows a pictorial representation of the network  where the nodes are colored
according to its degree: yellow = 1, green = 2, red = 3, blue=4, etc. The bottom panel shows the degree
distribution for two correlation thresholds. The inset depicts the degree distribution for an equivalent
randomly connected network. Data re-plotted from \cite{eguiluz}.}}
\end{figure}
To define the links, we denote as functionally connected those brain sites whose temporal evolution are
correlated beyond a positive pre-established value $r_c$, following an approach used previously in
\cite{dodel2002}. (Networks can be built as well by defining negative or positive correlations). Specifically,
we calculate the linear correlation coefficient between any pair of voxels, $x_1$ and $x_2$, as: \be r(x_1,
x_2)= \frac{\langle V(x_1,t) V(x_2,t)\rangle - \langle V(x_1,t)\rangle \langle V(x_2,t) \rangle} {\sigma
(V(x_1))\sigma (V(x_2))} ~, \ee where $\sigma^2(V(x)) = \langle V(x,t)^2\rangle - \langle V(x,t)\rangle^2$, and
$\langle \cdot \rangle$ represent temporal averages.

In Fig.~2, we show a typical network extracted with this technique for one subject in a finger tapping task
\cite{eguiluz}. The top panel shows the network's nodes (only a portion for illustration) colored according to
its degree and the bottom panel the degree distribution of the network. Degree is the mathematical term for each
voxel connectivity, being represented here as how many other voxels are temporally correlated with it. We find
that the degree distribution has a skewed distribution with a tail approaching a power law distribution with an
exponent around 2. As the threshold $r_c$ is decreased a maximum appears which shifts to the right as $r_c$ is
lowered. Other measures reveal that the number of links as a function of distance also decays as a power law.
This is so, from one voxel, the smallest scale able to be measured with this technique, to the largest, the size
of the brain. When we looked how the connectivity was arranged in the neighborhood of a node we found that
highly connected nodes were connected, on the average, with highly connected ones. This feature, only seen
before in social networks is inverse to what one expects from a hierarchical organization
\cite{maslov2003,newman2002,newman2003b}. We looked at two other statistical properties of these networks, path
length and clustering. The path length ($L$) between two voxels is the minimum number of links necessary to
connect both voxels. Clustering ($C$) is the fraction of connections between the topological neighbors of a
voxel with respect to the maximum possible. If voxel $i$ has degree $k_i$, then the maximum number of links
between the $k_i$ neighbors is $k_i(k_i-1)/2$. Thus, if $E_i$ is the number of links connecting the neighbors
then the clustering of voxel $i$, $C_i = 2E_i/ k_i(k_i-1)$. The average clustering of a network is given by $C =
1/N\sum_i C_i$, where $N$ is the number of voxels. The results are presented in Table 2 (average values for n =
22 datasets, $r_c=0.8$). From left to right are listed $N$, (number of nodes) $C$ (clustering coefficient), $L$
(shortest path length), the average degree $<k>$, and $\lambda$ the exponent of the degree distribution. The
clustering ($C_{rand}$) and path length ($L_{rand}$) values of an equivalent random network are also included
for comparison. In all cases, the coefficient $C$ remains four orders of magnitude larger than $C_{rand}$, the
clustering of a random network. This feature, together with the similarity of path length of the original
network and their randomized controls ($L$ and $L_{rand}$), is indicative of a small-world structure.

In this approach we define two voxels as ``linked" if they are temporally correlated beyond some value. This, of
course, does not mean they are mutually connected (via chemical or neuro-transmitter or anatomical paths). The
simplest counter-example is the case of a common input activating both sites. Further analysis will clarify if
this is the case for the networks we are studying. In this regard it is relevant to look at work done over the
last decade on the cortical connectivity (now ``linked" is used in the strong sense of being mutually connected)
by Sporns et al. \cite{sporns2000,sporns2003,spornschapter}, Hilgetag et al. \cite{hilgetag2000}, Young et al.
\cite{young93} and Scannel et al. \cite{scannel1999}. The analysis of connectivity matrix of these data sets
reveals that although the networks are highly clustered and exhibit relatively short path lengths (see Table 2)
as in small world networks, they exhibit a rather homogeneous degree distribution. In Figure 3 the macaque
cerebral cortex connectivity matrix \cite{young93} is plotted in the top panel. Because of the small statistics
(only 71 nodes), its degree distribution is computed as the cumulative density, plotted in the bottom panel of
the same figure. It can be seen that the degree distribution is not a power law, instead one sees that no area
has less than four links and that the majority are linked with about ten other areas.
\begin{table}
  \centering
\begin{tabular}{|c|c|c|c|c|c|c|c|}
   \hline
   $Network$& $N$ & $C$ & $L$ & $\langle k \rangle$ &$\lambda$& $C_{rand}$& $L_{rand}$  \\
   \hline
fMRI network & 4891 &   0.15 &   6.  & 4.12 & 2.2& 8.9$\times 10^{-4}$& 6.0 \\
   \hline
   Macaque~C.C.& 71 & 0.46 & 2.3 & 10.6 &  NA & 0.15 & 2.0 \\
   \hline
 \end{tabular}\\
  \caption{Statistical properties of human fMRI functional networks \cite{eguiluz} and macaque cerebral cortex
  connectivity \cite{spornschapter}.}
\end{table}
\begin{figure}[h!tb]
\centering \psfig{figure=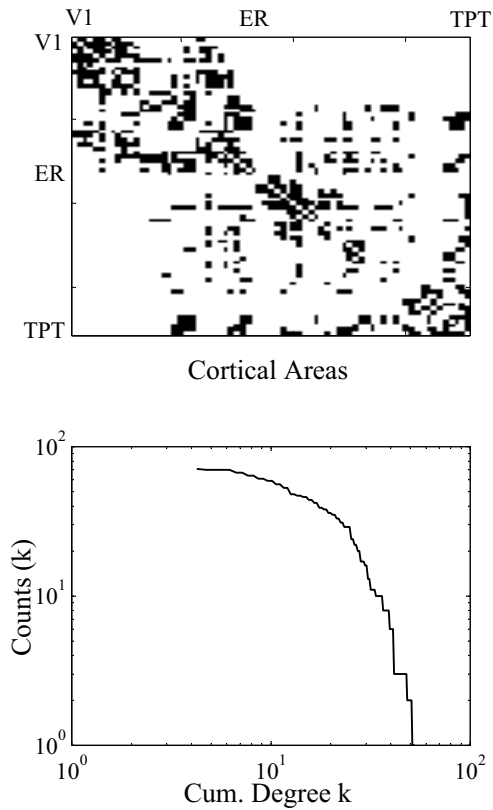,width=2.7 truein,clip=true,angle=0} \caption{\footnotesize{Top panel:
Connectivity matrix of the macaque cortex \cite{young93}. A black square denotes a connection between the 71
cortical areas: V1, V2, V3, VP, V3A, V4, VOT, V4T, MT, MSTD, MSTL, FST, PITD, PITV, CITD, CITV, AITD, AITV,
STPP, STPA, TF, TH, PO, PIP, LIP, VIP, DP, A7A, FEF, A46, TGV, ER, HIPP, A3A, A3B, A1, A2, A5, R1, S2, A7B, IG,
ID, A35, A4, A6, SMA, A3, A23, A24, A9, A32, A25, A10, A45, A12, A11, A13, G, PAAR, PAAL, PAAC, KA, PAL, PROA,
REIT, TGD, TS1, TS2, TS3, TPT (top to bottom, left to right). For reference are labelled visual (V1), enthorinal
(ER) and temporoparietal (TPT) cortices. Bottom panel: Cumulative degree distribution for the same data in the
top panel. In contrast with the data plotted in Fig.2, here there is no scale-free connectivity. Matrix data
from Ref. \cite{spornschapter} (available from website http://www.indiana.edu/$\sim$cortex/connectivity). } }
\end{figure}
The scale-free features illustrated in Fig. 2 reflects underlying long range correlations, i.e.,  brain activity
on a given area can be correlated with far away and apparently unrelated regions, something already documented
with other technology \cite{Linkenkaer2001}. In qualitative terms this means that, for instance, a concurrent
sound or simply imagery can influence thoughts or pain perception. Using the network approach here described,
various dynamical brain behavioral states can be studied in the future.
\section{Neuronal avalanches are critical.}
What are the neuronal mechanisms responsible for the correlations described in the previous section? From a
top-down approach, Varela was among the first to be concerned with the brain large scale dynamical properties
(reviewed in Ref. \cite{vanquyen}). Varela assumed that ``For every cognitive act, there is a singular and
specific large cell assembly that underlies its emergence and operation" \cite{varela95}. Efforts are underway
to formulate in neuronal terms this working hypothesis, also termed ``dynamic core" \cite{tononi1998}. Other
bottom-up complementary, approaches involve sophisticated recording in cortical structures with multielectrode
arrays (see a review in \cite{nicolelis2002}). A recent study \cite{plenz04} deserves special mention, because
it provides quantitative estimations of the dynamical properties and characteristic exponents opening the
possibility to model the neural mechanisms. Beggs and Plenz \cite{plenz04} studied the spontaneous neuronal
activity of cultured and acute slices of rat cortex using 60-channel multielectrode arrays. They documented that
during spontaneous activity the cortex typical activity shows intermittent avalanches. After computing the
statistics of several days worth of continuous cortical activity with a few millions of events, they showed that
the avalanche size distributions, expressed as total number of electrodes activated per run, demonstrates the
existence of a power law, with exponent $\alpha \sim 1.5$ as shown in Fig. 4. They also computed an average
branching activation ratio close to unity (calculated as the ratio between the current and future number of
excited electrodes). The authors showed that this branching ratio optimizes information transmission in
feedforward networks models. These aspects demonstrating criticality should be quantitatively accounted for by
futures theories of cortical dynamics.
\begin{figure}[h!tb]
\centering \psfig{figure=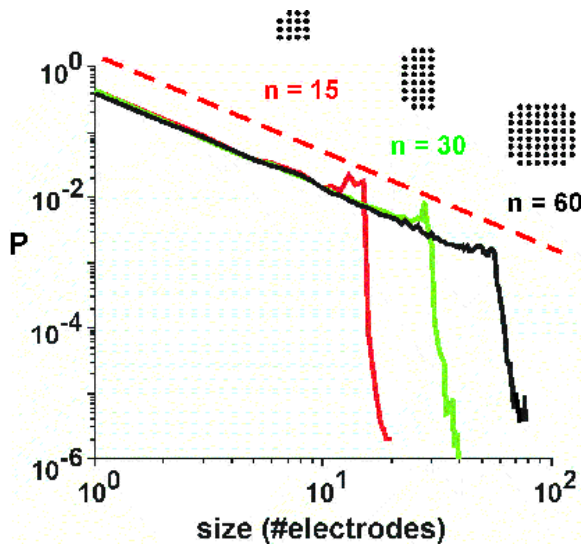,width=3 truein,clip=true,angle=0} \caption{\footnotesize{The size
distribution of neuronal avalanches in mature cortical cultured networks follows a power law with an exponent
$\sim 3/2$ (dashed line). The data, re-plotted from Figure 4 of \cite{plenz04}, shows the probability of
observing an avalanche covering a given number of electrodes for three sets of grid sizes shown in the insets
with n=15, 30 or 60 sensing electrodes (equally spaced at $200 \mu m$). The statistics is taken from data
collected from 7 cultures in recordings lasting a total of 70 hours and accumulating 58000 ($+-$ 55000)
avalanches per hour (mean $+-$ SD).}}
\end{figure}
\section{Per foresaw it.}
``Is biology too difficult for biologists?  And what can physics, dealing with the simple and lawful, contribute
to biology, which deals with the complex and diverse?". In such challenging terms Per Bak encouraged colleagues
to look at biology. Much before mainstream physics embraced biology, Per Bak was already convinced that ``The
big targets for physics theorists are biological evolution and the brain. These complex many-body problems might
have similarities to problems studied in particle- and solid-state physics." \cite{bak1}. In an attempt to shake
the stagnating state of these fields, by way of models and provocative metaphora, he suggested useful ways to
re-think the most important issues in these areas. At the same time, he often insisted, in his unforgiving way,
that mainstream ideas about neural models were an ``example of physicists leading a whole field astray", because
it was ``hard to imagine a biological foundation for the complicated procedures for updating the synaptic
strengths in those models" \cite{bak1}. It was fun, in a way, to witness him, a {\it physicist}, reminding {\it
biologists} and everybody else about the {\it biological} un-plausibility of current brain models and the need
to reconsider the constraints of self-organization as designing principle
\cite{chialvobak99,bakchialvo01,wakelingbak01,wakeling2003}. In his opinion, it was absolutely obvious that
self-organization is the driving mechanism designing Nature around us, regardless if it  was a human society,
millions of neurons or an ant swarm. It is, in this view, mandatory to understand first the general laws before
attempting to proceed with anything else. Per used to joke that, working out the further details will be just as
easy (and boring) as the ``cleaning after the party" . It is fascinating to see that a decade later,
self-organization issues are newsworthy even to engineering \cite{ottino}.

\vfill\eject


\begin{thebibliography}{99}
\bibitem{turing}A.M. Turing, Computing machines and intelligence. Mind {\bf 59}, 236 (1957).
\bibitem{bak0}P. Bak, How Nature works. Copernicus, 1998.
\bibitem{btw}P. Bak, C. Tang, and K. Wiesenfeld, Phys. Rev. Lett.~{\bf 59}, 381 (1987).
\bibitem{Paczuski04}M. Paczuski and D. Hughes. E-print arxiv.org Cond-mat/0311304.
\bibitem{millonas}E. Rauch, M.M. Millonas and D.R. Chialvo, Phys. Lett. A. ~{\bf 207}, 185 (1995).
\bibitem{chialvo90}D.R. Chialvo, R.F. Gilmour RF, and J. Jalife, Nature {\bf 343}, 653 (1990).
\bibitem{davidenko90}J. Davidenko, P. Kent, D.R. Chialvo, D. Michaels and J. Jalife, Proc. Natl. Acad. Sci. USA {\bf 87} 8785 (1990).
\bibitem{albert2000}R. Albert and  A.-L Barabasi, Rev. Mod. Phys. {\bf 74}, 47 (2002).
\bibitem{newman2003a}M.E.J. Newman, SIAM Review {\bf 45}, 167 (2003).
\bibitem{jasny2003}B.R. Jasny and L.B. Ray, Science {\bf 301}, 1863 (2003).
\bibitem{doro2002}S.N. Dorogovtsev and J.F.F. Mendes,  Adv. Phys. {\bf 51}, 1079 (2002).
\bibitem{strogatz2001}S.H. Strogatz, Nature {\bf 410}, 268 (2001).
\bibitem{watts1998}D.J. Watts  and   S.H. Strogatz, Nature {\bf 393}, 440 (1998).
\bibitem{dodel2002}S. Dodel, J.M. Herrmann  and  T. Geisel, Neurocomputing {\bf 44}, 1065 (2002).
\bibitem{eguiluz}V.M. Eguiluz, D.R. Chialvo, G. Cecchi, M. Baliki and V. Apkarian, E-print arxiv.org  Cond-mat/0309092.
\bibitem{maslov2003}S. Maslov, K. Sneppen, and  U. Alon,  In Handbook of graphs and networks.
From the genoma to the internet (Eds. S. Bornholdt and H.G. Schuster) 169-198. (Wiley -VCH  and Co. Weinheim,
2003).
\bibitem{newman2002}M.E.J. Newman,  Phys. Rev. Lett. {\bf 89}, 208701 (2002).
\bibitem{newman2003b}M.E.J. Newman and J. Park, Phys. Rev. E {\bf 68}, 036122 (2003).
\bibitem{sporns2000}O. Sporns, G. Tononi and  G.M. Edelman, Cerebral Cortex {\bf 10}, 127 (2000).
\bibitem{sporns2003}O. Sporns and  G. Tononi, Complexity {\bf 7}, 28 (2003).
\bibitem{spornschapter}O. Sporns, Graph theory methods for the analysis of neural connectivity patterns.
In Neuroscience Databases. A Practical Guide. (Ed. R. Kotter) 169-183. (Kluwer Pub., Boston, MA, 2002).
\bibitem{hilgetag2000}C.C. Hilgetag, G.A.P.C. Burns, M.A. O'Neill, J.W. Scannell and  M.P. Young, Phil. Trans. R. Soc. Lond. B {\bf 355}, 91 (2000).
\bibitem{young93}M.P. Young, Proceedings of the Royal Society London B {\bf 252},13 (1993).
\bibitem{scannel1999}J.W. Scannell, G.A.P.C. Burns, C.C. Hilgetag, M.A. O'Neil and M.P. Young, Cerebral Cortex {\bf 9}, 277 (1999).
\bibitem{Linkenkaer2001}K. Linkenkaer-Hansen, V.V. Nikouline, J.M. Palva and R.J. Ilmoniemi, J. Neuroscience {\bf 21}, 1370 (2001).
\bibitem{vanquyen}M. Le Van Quyen, Biol. Res. {\bf 36}, 67 (2003).
\bibitem{varela95}F.J. Varela, Biol. Res. {\bf 28},81 (1995).
\bibitem{tononi1998}G. Tononi and G.M. Edelman, Science {\bf 282}, 1846 (1998).
\bibitem{nicolelis2002} M.A.L. Nicolelis and S. Ribeiro, Curr. Opin. Neurobiol. {\bf 12}, 602 (2002).
\bibitem{plenz04}J.M. Beggs and D. Plenz, J. Neuroscience {\bf 23}, 11167 (2003).
\bibitem{bak1}P.~Bak, Nature {\bf 391}, 652 (1998).
\bibitem{chialvobak99}D.R. Chialvo and P. Bak, Neuroscience {\bf 90}, 1137 (1999).
\bibitem{bakchialvo01}P. Bak and D.R. Chialvo, Phys. Rev. E  {\bf 63}, 031912 (2001).
\bibitem{wakelingbak01}J. Wakeling and P. Bak, Phys. Rev. E. {\bf 64}, 051920 (2001).
\bibitem{wakeling2003}J. Wakeling, Physica A. {\bf 325}, 561 (2003).
\bibitem{ottino}J.M. Ottino, Nature {\bf 427}, 399 (2004).
\end{thebibliography}
\end{document}